\documentclass[11pt]{article}
\usepackage[dvips]{graphics}
\newcommand{\apm}{\!\stackrel{\leftrightarrow\;}{\partial_{\mu}}\!}
\newcommand{\apnablal}{\!\stackrel{\leftrightarrow\;}{\nabla_j}\!}
\newcommand{\aptl}{\!\stackrel{\leftrightarrow\;}{\partial_{t_j}}\!}
\newcommand{\apnul}{\!\stackrel{\leftrightarrow\;}{\partial_0}\!}
\begin{document}

\noindent
\Large
{\bf BOHMIAN PARTICLE TRAJECTORIES IN RELATIVISTIC BOSONIC 
QUANTUM FIELD THEORY}
\normalsize
\vspace*{1cm}

\noindent
{\bf Hrvoje Nikoli\'c}

\vspace*{0.5cm}
\noindent
{\it 
Theoretical Physics Division \\
Rudjer Bo\v{s}kovi\'{c} Institute \\
P.O.B. 180, HR-10002 Zagreb, Croatia \\
E-mail: hrvoje@thphys.irb.hr}

\vspace*{2cm}

\noindent
We study the de Broglie--Bohm interpretation of
bosonic relativistic quantum mechanics and argue
that the negative densities and superluminal
velocities that appear in this interpretation
do not lead to inconsistencies. After that,
we study particle trajectories in bosonic quantum
field theory.
A new continuously changing hidden variable -
the {\it effectivity} of a particle (a number
between 0 and 1) - is postulated. This variable 
leads to a causal description of processes
of particle creation and destruction.
When the field enters one of
nonoverlapping wave-functional packets
with a definite number of particles, then the effectivity of
the particles corresponding to this packet becomes equal to 1,
while that of all other particles becomes equal to 0.
\vspace*{0.5cm}


\section{INTRODUCTION}

The de Broglie--Bohm (dBB) interpretation of 
nonrelativistic quantum mechanics (QM) and 
relativistic quantum field theory (QFT)
\cite{bohm,bohmPR1,bohmPR2,holPR,holbook}
offers a clear answer (see, e.g., Refs.~\cite{bohmPRL,bohmN,dewN}) 
to notoriously difficult ontological questions that 
arise from the conventional interpretation of QM and QFT.
Yet, the current form of the 
dBB theory of motion is still not completely satisfactory.  
Relativistic wave equations of bosonic particles lead to 
superluminal velocities and motions backwards in time 
\cite{cuf,kyp,dew}. Similarly, the particle density of relativistic 
bosonic fields may be negative \cite{holPR,holbook}. 
There exist formal ways to overcome these problems 
by linearizing the Klein-Gordon equation \cite{holNC} or by
using the energy-momentum tensor to define timelike particle 
trajectories \cite{hor}, but the resulting particle 
densities are not in correspondence 
with the conventional notion of particle in QFT
\cite{holbook}. This led to a 
conclusion that bosons do not have particle trajectories, i.e., that 
bosons are causally evolving fields \cite{holPR,holbook}. 
Thus, in contrast with the conventional QFT, according to 
the current 
version of the dBB theory of motion there is a great asymmetry between 
bosons and fermions because relativistic fermions do have 
particle trajectories and are not described by quantum fields 
\cite{bohmPR2,dew,holbook}. 

One of the arguments in favor of 
the nonexistence of particle trajectories for bosons is a difficulty 
with a causal description of trajectories for particles that are created 
or destroyed or for states with an indefinite 
number of particles \cite{holbook}. However, the same problem remains
for fermion trajectories as well, so the current 
version of the dBB theory of fermions cannot describe observed effects of 
fermion creation and destruction. There is also an alternative 
version of the dBB theory of fermions that describes creation and
destruction of fermions \cite{holfer}, but this version does not 
incorporate particle trajectories.
It has been argued that  
processes of particle creation and destruction cannot be described in a 
deterministic way \cite{sud,durr}. There is an attempt 
to describe that in a deterministic way by introducing 
a direct particle interaction \cite{shoj}, 
but it does not seem that it leads to the same statistical predictions 
as the conventional QFT. 
A recent version of the dBB theory of fermions 
\cite{col}, based on earlier work \cite{bell1,bell2},
describes creation and
destruction of fermions and incorporates particle trajectories.

In this paper we propose a solution of the problems with the 
dBB interpretation of relativistic bosons discussed above. 
We argue that superluminal velocities, motions backwards 
in time and negative densities do not lead to any inconsistencies. 
Moreover, it seems that these properties are desirable for
a causal description of some QFT effects.
After that, we propose a causal interpretation of multiparticle 
wave functions that result from QFT. In this 
interpretation, all particles that may exist in a state with an 
indefinite number of particles do actually exist for all 
time. Particles are never 
really created or destructed. However, to each particle, 
we attribute a new 
deterministic continuously evolving nonlocal hidden variable - the 
{\it effectivity} $e$ of a particle. A particle with $e=0$ has
the effects as if it did not exist, while that with $e=1$ has
the effects as a particle in the usual sense. We explain how in 
the process of measurement all effectivities take values 
equal to either 0 or 1, which has the same effect as if the 
wave functional had ``collapsed" to a state with a definite number 
of particles.   

The paper is organized as follows. In Sec.~\ref{sec2} we study 
the Bohmian particle trajectories in ``one-particle" relativistic 
QM. In
Sec.~\ref{sec3} we study many-particle wave 
functions in interacting QFT and give a 
causal interpretation of them in terms of Bohmian particle trajectories.
A critical discussion of our results
is given in Sec.~\ref{sec4}.
In the Appendix, we present a short review of the general theory 
of quantum measurements in the dBB interpretation. 

\section{PARTICLE TRAJECTORIES IN RELATIVISTIC QM}
\label{sec2}

\subsection{Basic Equations}

Consider a {\em real} scalar field $\phi(x)$ satisfying the Klein-Gordon 
equation (in a Minkowski metric 
$\eta_{\mu\nu}\!=\!{\rm diag} (1,-1,-1,-1)$)
\begin{equation}\label{KG}
(\partial_0^2-\nabla^2+m^2)\phi=0.
\end{equation}
Let $\psi\! =\!\phi^+$ ($\psi^*\! =\!\phi^-$) be the positive (negative) 
frequency 
part of the field $\phi=\phi^+ + \phi^-$. The particle current is 
\cite{nikol_PLB,nikol_long,nikol}
\begin{equation}\label{loc2}
j_{\mu}=i\psi^*\apm\psi ,
\end{equation}
where $A\apm B\equiv A\partial_{\mu}B-B\partial_{\mu}A$.
The quantity
\begin{equation}\label{N}
N=\int d^3x\, j_0
\end{equation}
represents the positive-definite number of particles (not the charge!).
This is most easily seen from the plane-wave expansion
$\phi^+(x)=\int d^3k\, a({\bf k})e^{-ikx}/\sqrt{(2\pi)^3 2k_0}$, 
because then
$N=\int d^3k\, a^{\dagger}({\bf k})a({\bf k})$. (For more details, see
Refs.~\cite{nikol_PLB,nikol_long,nikol},
where it is shown that the particle current and the decomposition 
$\phi=\phi^+ + \phi^-$ makes sense even when a background 
gravitational field or some other potential is present.)
The particle density $j_0$ can also be written as 
$j_0=i(\phi^-\pi^+ - \phi^+\pi^-)$ (where $\pi=\pi^+ + \pi^-$ is the 
canonical momentum), which is the form used in Refs.~\cite{holPR,holbook}.

Alternatively, $\phi$ may be interpreted not as a field containing 
an arbitrary number of particles, but rather as a 
one-particle wave function. Historically, this later interpretation 
was attempted before the former one. Contrary 
to a field, a wave function is not an observable. 
In this later interpretation, 
which we employ in the rest of this section, 
it is convenient to normalize 
the wave function $\phi$ such that $N=1$. 

The current (\ref{loc2}) is conserved:
\begin{equation}\label{conserv}
\partial_{\mu}j^{\mu}=0,
\end{equation}
which implies that (\ref{N}) is also conserved: $dN/dt=0$.
In the causal interpretation, we postulate that the particle has the 
trajectory determined by
\begin{equation}\label{traj1}
\frac{dx^{\mu}}{d\tau}=\frac{j^{\mu}}{2m\psi^*\psi}.
\end{equation}  
The affine parameter $\tau$ may be eliminated by writing the equation
for the trajectory as
\begin{equation}\label{traj3}
\frac{d{\bf x}}{dt}=\frac{{\bf j}(t,{\bf x})}{j_0(t,{\bf x})},
\end{equation}
where $t=x^0$, ${\bf x}=(x^1,x^2,x^3)$, ${\bf j}=(j^1,j^2,j^3)$.

By writing $\psi=Re^{iS}$, where $R$ and $S$ are real functions,
we can also write all the equations above in the Hamilton--Jacobi form.
Eq.~(\ref{traj1}) can be written as
\begin{equation}\label{traj2}
\frac{dx^{\mu}}{d\tau}=-\frac{1}{m}\partial^{\mu}S.
\end{equation}
The Klein-Gordon equation (\ref{KG}) is equivalent to a set of
two equations
\begin{equation}\label{cont}
\partial^{\mu}(R^2\partial_{\mu}S)=0,
\end{equation}
\begin{equation}\label{HJ}
-\frac{(\partial^{\mu}S)(\partial_{\mu}S)}{2m} +\frac{m}{2} +Q=0,
\end{equation}
where (\ref{cont}) is the conservation equation
(\ref{conserv}), (\ref{HJ}) is the
Hamilton--Jacobi equation, and
\begin{equation}
Q=\frac{1}{2m}\frac{\partial^{\mu}\partial_{\mu}R}{R}
\end{equation}
is the quantum potential. From (\ref{traj2}), (\ref{HJ}),
and the identity
\begin{equation}
\frac{d}{d\tau}=\frac{dx^{\mu}}{d\tau}\partial_{\mu},
\end{equation}
we find the equation of motion
\begin{equation}
m\frac{d^2x^{\mu}}{d\tau^2}=\partial^{\mu}Q.
\end{equation}

It is easy to show that all the equations above have the correct
nonrelativistic limit. In particular, by writing
\begin{equation}
\psi=\frac{e^{-imt}}{\sqrt{2m}}\chi
\end{equation}
and using $|\partial_t\chi|\ll m|\chi|$,
$|\partial^2_t\chi|\ll m|\partial_t\chi|$,
from (\ref{loc2}) and (\ref{KG}) we find the approximate
equations
\begin{equation}
j_0=\chi^*\chi,
\end{equation}
\begin{equation}\label{nonrel}
-\frac{\nabla^2}{2m}\chi=i\partial_t\chi,
\end{equation}
which are the usual nonrelativistic equations for the conserved
probability density and for the evolution of the wave function
$\chi$, respectively.

When the nonrelativistic limit cannot be applied, then the quantity 
$j_0$ is not positive 
definite, so in that case it cannot be interpreted as a probability 
density. For such a general situation, we give the 
interpretation of $j_0$ in the following subsection.

\subsection{Physical Interpretation}

\begin{figure}[h]
\centerline{\includegraphics{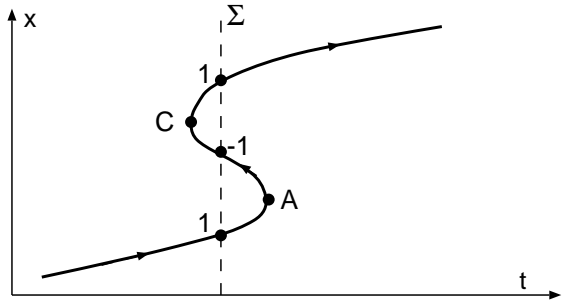}}
\caption{A part of a boson-particle trajectory. The particle
moves backwards in time from A to C. The dashed line
represents a spacelike hypersurface $\Sigma$ intersected by the trajectory
at 3 points. The particle density $j_0$ is positive at the points
marked by 1 and negative at the point marked by $-1$.
The number of particles at $\Sigma$ is equal to 1
because $1+(-1)+1=1$. In the reinterpretation of
negative densities, all particles move forwards in time and the
physical number of particles at $\Sigma$ is equal to 3.
A pair of particles is created at C and annihilated at A. At these
two points, $j_0=0$.}
\end{figure}

A typical trajectory that may arise as a solution of (\ref{traj3}) 
is sketched in Fig.~1. The velocity of a particle may be 
superluminal. Moreover, the particle may move backwards in time, 
which occurs at the points where $j_0<0$. Of course, superluminal 
velocities and negative densities are closely related, because any 
vector $u_{\mu}$ with $u_0>0$ may be transformed by a 
Lorentz transformation to $u'_{\mu}$ with $u'_0<0$, if and only if 
$u_{\mu}$ is spacelike. The 3-velocity is infinite 
at the points where $j_0=0$ and ${\bf j}\neq 0$.

The motion backwards in time does not lead to causal paradoxes 
when one realizes that it is physically indistinguishable from 
a motion forwards in time but with negative 
energy \cite{cost}. Therefore, it is 
natural to reinterpret the negative densities by introducing the 
physical number of particles
\begin{equation}\label{Np}
N_{{\rm phys}}=\int d^3x\, |j_0|,
\end{equation}
where the physical particle density $|j_0|$ is {\em nonnegative}.
Contrary to (\ref{N}), the physical number of particles is not conserved. 
A pair of particles, one with positive and the other with negative 
energy, may be created at one point and annihilated at another point. 
(Recall that 
this is not a particle-antiparticle pair because we are studying a 
real field $\phi$). Needless to say, this creation and 
annihilation of highly off-shell particles resembles the behavior 
of ``virtual" particles in the conventional QFT. 

Note that superluminal velocities do not contradict existing 
experiments because, according to the general theory of 
quantum measurements \cite{bohm,bohmPR1,holbook} 
(see also the Appendix),
the outcome of an ideal measurement of the 
4-momentum may only be an on-shell eigenvalue $p_{\mu}$ 
attributed to the eigenfunction $\psi_p(x)=\exp(-ip_{\mu}x^{\mu})$ that 
solves (\ref{KG}). This is because the total final wave function that 
describes the entanglement between the measured particle and 
the measuring apparatus during an   
ideal quantum measurement can be written as
\begin{equation}\label{jed17}
\psi(x,y)=\int d^3p\: c_p\psi_p(x)\chi_p(y),
\end{equation}
where $y$ denotes the coordinates of the measuring apparatus. 
The wave functions $\chi_p(y)$ do not overlap, in the sense 
that $\chi_p(y)\chi_{p'}(y)=0$ for $p\neq p'$. Therefore, as 
explained in Refs.~\cite{bohm,bohmPR1,holbook}
and the Appendix, the wave functions 
$\psi_p(x)\chi_p(y)$ constitute a set of nonoverlaping 
``channels",
so a particle in a ``channel" behaves as if the other ``channels"  
did not exist. 
 
The creation and annihilation 
of free particles described above does not occur 
during an ideal measurement of a particle momentum because the 
trajectories are straight lines when $\psi(x)=\psi_p(x)$.
Similarly, when a position of a particle is measured with  
good accuracy, then $\psi_p(x)$ are replaced by wave 
functions well localized in space, so multiple particles 
created during the measurement are 
confined inside a volume too small to allow the 
experimental distinction 
between the particles at different positions.
This explains why the creation and annihilation of free 
particles is not seen in experiments. 
However, in principle, 
it would not be impossible to see such creations and annihilations if   
we knew how to perform measurements radically different from the 
ideal ones. Thus, our theory leads to  
predictions that differ from those of the conventional 
approach and might be observable in future.

There are also no causal paradoxes because 
the trajectories of particles, including those that are created 
or annihilated, are uniquely and self-consistently determined by 
specifying the fields $\phi({\bf x})$ and $\pi({\bf x})$ at an initial 
spacelike Cauchy hypersurface and one initial position of the particle 
at that hypersurface. It may appear, as in Fig.~1, that 
the specification of one particle position implies the 
simultaneous existence 
of other particles at different positions, but 
this nonlocal feature does not contradict the causality in the 
sense above.

Finally, note that the existence of motions backwards in time, 
which may be reinterpreted as 
motions with negative energy, may be regarded as a desirable property, 
because the black-hole evaporation is often viewed 
as a process in which a pair of particles is produced 
at the horizon, so that the 
particle with positive energy escapes from the horizon, while the particle 
with negative energy is absorbed by the black hole \cite{bd}. However, 
in this paper, we do not further explore the causal description of 
a specific process of particle creation such as the black-hole evaporation. 
Instead, we give a general formalism that describes the particle creation 
in the dBB interpretation of interacting QFT.  

\section{PARTICLE TRAJECTORIES IN RELATIVISTIC QFT}
\label{sec3}

\subsection{Wave Functionals and Many-Particle Wave Functions}

In the Heisenberg picture, the hermitian field operator $\hat{\phi}(x)$ 
satisfies the equation of motion
\begin{equation}\label{KGq}
(\partial_0^2-\nabla^2+m^2)\hat{\phi}=J(\hat{\phi}),
\end{equation}
where $J(\hat{\phi})$ is a nonlinear function that describes the 
interaction. In a more general case, $J$ may also be a function of 
other quantum fields or a function of background classical fields.
In the Schr\"odinger picture, the time evolution is determined by the 
Schr\"odinger equation
\begin{equation}\label{sch}
H[\phi,-i\delta/\delta\phi]\Psi[\phi,t]=i\partial_t\Psi[\phi,t], 
\end{equation}
where $\Psi[\phi,t]$ is a functional with respect to $\phi({\bf x})$ 
and a function with respect to $t$.
A normalized solution of (\ref{sch}) can be expanded as
\begin{equation}\label{exp}
\Psi[\phi,t]=\sum_{n=0}^{\infty}\tilde{\Psi}_n[\phi,t],
\end{equation}
where $\tilde{\Psi}_n$ are unnormalized $n$-particle wave functionals.
Since any (well-behaved) function $\phi({\bf x})$ can be 
Fourier expanded, the functionals $\tilde{\Psi}_n$
can be further expanded as \cite{holbook,long} 
\begin{equation}\label{exp2}
\tilde{\Psi}_n[\phi,t]=\int d^3k_1\cdots d^3k_n\,
c_n(\vec{{\bf k}}^{(n)},t) \Psi_{n,\vec{\bf k}^{(n)}}[\phi],
\end{equation}
where $\vec{\bf k}^{(n)}=\{ {\bf k}_1,\ldots,{\bf k}_n\}$. 
The functionals $\Psi_{n,\vec{\bf k}^{(n)}}[\phi]$ constitute 
a complete orthonormal basis for the expansion of an arbitrary 
functional $\Psi[\phi]$. 
(This basis generalizes the complete
orthonormal basis consisting of the Hermite functions
$
h_n(x)\! =\!(\sqrt{\pi}2^n n!)^{-1/2} e^{-x^2/2} H_n(x),
$
where $H_n(x)$ are the Hermite polynomials. For more details, see
Ref.~\cite{long}).
They have a property
\begin{equation}\label{=0}
\int{\cal D}\phi\, \Psi_0^*[\phi] \phi({\bf x}_1)\cdots\phi({\bf x}_{n'})
\Psi_{n,\vec{\bf k}^{(n)}}[\phi] =0 \;\; {\rm for} \;\; 
n'\neq n.
\end{equation} 

For free fields, i.e., when $J=0$ in (\ref{KGq}), 
the coefficients $c_n(\vec{{\bf k}}^{(n)},t)$ 
have a simple oscillating behavior of the form
\begin{equation}
c_n(\vec{{\bf k}}^{(n)},t)=c_n(\vec{{\bf k}}^{(n)})
e^{-i\omega_n(\vec{{\bf k}}^{(n)}) t},
\end{equation}
where
\begin{equation}
\omega_n(\vec{{\bf k}}^{(n)})=E_0+\sum_{j=1}^n 
\sqrt{{\bf k}_j^2+m^2}
\end{equation}
and $E_0$ is the vacuum energy. 
In this case, the quantities $|c_n(\vec{{\bf k}}^{(n)},t)|^2$ do not 
depend on time, which means that the number of particles 
(corresponding to the quantized version of (\ref{N})) is conserved. 
In a general case with interactions, the Schr\"odinger 
equation (\ref{sch}) leads to a more complicated time dependence 
of the coefficients $c_n(\vec{{\bf k}}^{(n)},t)$, so the number of 
particles is not conserved. 

For free fields, the (unnormalized) $n$-particle 
wave function is \cite{schweber}
\begin{equation}\label{wf1} 
\psi_n(\vec{{\bf x}}^{(n)},t)=
\langle 0|\hat{\phi}(t,{\bf x}_1)\cdots\hat{\phi}(t,{\bf x}_n)
|\Psi\rangle ,
\end{equation}
where $\vec{{\bf x}}^{(n)}=\{ {\bf x}_1,\ldots,{\bf x}_n\}$. 
(The multiplication of the right-hand side of (\ref{wf1}) by 
$(n!)^{-1/2}$ would lead to a normalized wave function only if 
$\Psi\! =\!\tilde{\Psi}_n$ in (\ref{exp}).)
The generalization of (\ref{wf1}) to the interacting 
case is not trivial because, in systems with an unstable 
vacuum, it is not obvious what the analog of the state $\langle 0|$ 
in (\ref{wf1}) is. To treat this problem correctly, we find 
the Schr\"odinger picture more convenient.
Using the Schr\"odinger picture, (\ref{wf1}) becomes
\begin{equation}\label{wf2}
\psi_n(\vec{{\bf x}}^{(n)},t)=
\int\!{\cal D}\phi\, \Psi_0^*[\phi] e^{-i\varphi_0(t)}
\phi({\bf x}_1)\cdots\phi({\bf x}_n) \Psi[\phi,t],
\end{equation}
where $\varphi_0(t)=-E_0 t$. For the interacting case, we define 
the wave function to be given by (\ref{wf2}), but with a 
different phase $\varphi_0(t)$. This phase is defined by an expansion
of the form of (\ref{exp}):
\begin{equation} 
\hat{U}(t)\Psi_0[\phi]=r_0(t)e^{i\varphi_0(t)}\Psi_0[\phi] 
+\sum_{n=1}^{\infty}\ldots ,
\end{equation}
where $r_0(t)\!\geq\! 0$ and 
$\hat{U}(t)\! =\! U[\phi,-i\delta/\delta\phi, t]$ 
is the unitary time-evolution operator. 
From (\ref{exp}), (\ref{exp2}), and (\ref{=0}) we see that, 
even in the interacting case, only 
the $\tilde{\Psi}_n$-part of $\Psi$ contributes to (\ref{wf2}),  
which justifies to call $\tilde{\Psi}_n$ the $n$-particle wave functional. 

The wave function (\ref{wf1}) can also be generalized to a
nonequal-time wave function
\begin{equation}\label{wf3}
\psi_n(\vec{x}^{(n)})=S_{\{ x_j\} }
\langle 0|\hat{\phi}(x_1)\cdots\hat{\phi}(x_n)
|\Psi\rangle .
\end{equation}
Here $S_{\{ x_j\} }$ denotes symmetrization over all $x_j$, which is
needed because the field operators do not commute for nonequal times.
For the interacting case, the nonequal-time wave function
is defined as a generalization of
(\ref{wf2}) with the replacements
\begin{eqnarray}
& \displaystyle
\phi({\bf x}_j)\rightarrow 
\hat{U}^{\dagger}(t_j)\phi({\bf x}_j)\hat{U}(t_j), &
\nonumber \\
& \Psi[\phi,t]\rightarrow \hat{U}^{\dagger}(t) \Psi[\phi,t] =\Psi[\phi],
\nonumber \\
& e^{-i\varphi_0(t)}\rightarrow e^{-i\varphi_0(t_1)} \hat{U}(t_1), &
\end{eqnarray}
followed by symmetrization.

For free fields, the wave function (\ref{wf3}) satisfies the
equation
\begin{equation}\label{wfeq}
\sum_{j=1}^{n} [(\partial^{\mu}\partial_{\mu})_j +m^2]
\psi_n(\vec{x}^{(n)})=0.
\end{equation}   
Taking the nonrelativistic limit first and then putting
$t_1=\cdots=t_n=t$ in (\ref{wfeq}),
we find the multiparticle generalization of (\ref{nonrel})
\begin{equation} 
\sum_{j=1}^{n} \frac{-\nabla^2_j}{2m}\, \chi_n(\vec{{\bf x}}^{(n)},t)
=i\partial_t \chi_n(\vec{{\bf x}}^{(n)},t).
\end{equation}
This is the standard nonrelativistic multiparticle Schr\"odinger
equation, usually postulated without reference to QFT.

\subsection{Causal Interpretation}

In the dBB interpretation, the field $\phi(x)$ has a causal evolution
determined by \cite{holPR,holbook}
\begin{equation}\label{KGc}
(\partial_0^2-\nabla^2+m^2)\phi(x)=J(\phi(x))
-\left(\frac{\delta Q[\phi,t]}{\delta\phi({\bf x})}
\right)_{\phi({\bf x})=\phi(x)} ,
\end{equation}   
where
\begin{equation} 
Q=-\frac{1}{2|\Psi|} \int d^3x \frac{\delta^2|\Psi|}
{\delta\phi^2({\bf x})}
\end{equation}   
is the quantum potential.
However, the $n$ particles 
attributed to the wave function $\psi_n$ also have causal trajectories. 
They are determined by a generalization of (\ref{traj3}) as  
\begin{equation}\label{traj3l}
\frac{d{\bf x}_{n,j}}{dt}=\left( \frac{\psi^*_n(\vec{x}^{(n)})
\apnablal \psi_n(\vec{x}^{(n)})} {\psi^*_n(\vec{x}^{(n)}) 
\aptl \psi_n(\vec{x}^{(n)})} \right)_{t_1=\cdots =t_n=t},
\end{equation}
for $j=1,\ldots,n$. 
The norms of the wave functions $\psi_n$ change with time because  
the coefficients $|c_n|$ change with time. 
However, in (\ref{traj3l}), the norms are irrelevant.
Even when $c_n\rightarrow 0$ (for some, but not all $n$) during the 
evolution (for example, this may occur for $t\rightarrow\pm\infty$ in a 
scattering process), 
the ratio on the right-hand side of (\ref{traj3l}) is well defined. 
Of course, there may exist isolated points at which 
the ratio diverges, but we already know how to physically 
interpret these points as points at which the velocity is 
infinite.
This means that {\em these $n$-particles have well-defined trajectories 
even when the probability (in the conventional interpretation of 
QFT) of their experimental detection is equal to zero}. 
In the dBB interpretation of QFT, we can introduce 
a new causally evolving parameter $e_n[\phi,t]$ defined as 
\begin{equation}\label{en}
e_n[\phi,t]=\frac{|\tilde{\Psi}_n[\phi,t]|^2}
{\displaystyle\sum_{n'=0}^{\infty}|\tilde{\Psi}_{n'}[\phi,t]|^2}.
\end{equation} 
The evolution of this parameter is determined by the evolution 
of $\phi$ given by 
(\ref{KGc}) and by the solution (\ref{exp}) of (\ref{sch}), so one does 
not need a separate evolution equation for $e_n[\phi,t]$.
This parameter might be interpreted as a probability that there are $n$ 
particles in the system at the time $t$ if the field is equal 
(but not measured!) to 
$\phi({\bf x})$ at that time. However, in the dBB interpretation, we do 
not 
want an intrinsically stochastic interpretation. Therefore, we postulate
that $e_n$ is an {\em actual} property of the particles guided by the 
wave function $\psi_n$. We call this property the {\em effectivity} of these 
$n$ particles. From the point of view of the conventional interpretation 
of QFT, this is a nonlocal hidden variable attributed 
to the particles. We introduce this parameter in order to provide a 
deterministic description of the creation and destruction of particles.
We postulate that the effective mass of a particle 
guided by $\psi_n$ is
$m_{{\rm eff}}=e_n m$, 
and similarly for the energy,
momentum, charge, and other measurable quantities that are proportional 
to the number of particles. This is achieved by postulating 
that the mass density $\rho_{{\rm mass}}$ is given by 
\begin{equation}\label{goldstein}
\rho_{{\rm mass}}({\bf x},t)=m\sum_{n=1}^{\infty}e_n
\sum_{j=1}^{n} \delta^3({\bf x}-{\bf x}_{n,j}(t)),
\end{equation} 
and similarly for the other quantities.
Therefore, if $e_n=0$, then these $n$ particles 
are ineffective, i.e., their effect is as if they did not exist. Similarly, 
if $e_n=1$, then their effect is as they exist in the usual sense.    
However, since the trajectories are defined even for the particles for 
which $e_n=0$, the initial condition for particle positions contains 
one initial condition for the particle guided by $\psi_1$, 
two initial conditions for the particles guided by $\psi_2$, and so on, 
which leads to an {\em infinite} number of initial positions.     
In this way, QFT is really a theory of an infinite number of particles, 
although some of them may be ineffective. (This resembles the conventional
picture  
of QFT as a theory of an infinite number of particles, although some of them
may be ``virtual".)

The formalism may also be generalized to a case with many different 
particle species described by various bosonic fields. 
The wave functional
$\tilde{\Psi}_n$ generalizes to $\tilde{\Psi}_{\{ n\} }$, where 
$\{ n\}=\{ n_1,\ldots,n_{N_s}\}$ and $N_s$ is the number of different particle 
species. 
Equation (\ref{en}) generalizes to  
\begin{equation}\label{engen}
e_{\{ n\} }[\{\phi\},t]=\frac{|\tilde{\Psi}_{\{ n\} }[\{\phi\},t]|^2}
{\displaystyle\sum_{\{ n'\} }|\tilde{\Psi}_{\{ n'\} }[\{\phi\},t]|^2},
\end{equation}
where $\{\phi\}=\{\phi_1,\ldots,\phi_{N_s}\}$.

In experiments in which the number of particles is measured, one finds 
that a particle either exists or does not exist. In other words, the measured 
effectivity is either 0 or 1. At first sight, this is in contradiction with 
our theory that allows effectivities 
to take {\em any} value from the compact interval $[0,1]$. However, 
there is no contradiction! If different $\tilde{\Psi}_n$'s in the expansion 
(\ref{exp}) do not overlap in the $\phi$ space, then these
$\tilde{\Psi}_n$'s constitute a set of nonoverlapping ``channels" 
for the causally evolving field $\phi$. The field necessarily 
enters {\em one and only one} of the ``channels". From (\ref{en}) we see 
that $e_n=1$ for the ``channel" $\tilde{\Psi}_n$ that is not empty, while 
$e_{n'}=0$ for all other empty ``channels" $\tilde{\Psi}_{n'}$.
(This is because, owing to the assumption that different 
$\tilde{\Psi}_n$'s do not overlap, $\tilde{\Psi}_{n'}=0$ at
the configuration
$\phi$ which is from the support of $\tilde{\Psi}_n$.)
The effect is the same as if the wave functional $\Psi$ ``collapsed" into 
one of the states $\tilde{\Psi}_n$ with a definite number of particles. 

In a more general situation, different $\tilde{\Psi}_n$'s of the measured 
particles may overlap. However, the general theory of ideal quantum 
measurements \cite{bohm,bohmPR1,holbook} provides that the total wave 
functional can be written again as a sum of nonoverlapping wave 
functionals
in the $\{\phi\}$ space, where one of the fields represents the measured 
field, while the others represent the fields of 
the measuring apparatus. In this general case, one and only one of
$\tilde{\Psi}_{\{ n\}} $'s in (\ref{engen}) becomes nonempty, so  
the corresponding $e_{\{ n\}} $ becomes equal to 1, while all other 
$e_{\{ n'\} }$'s become equal to 0.

The essential point is that, from the point of view of an 
observer who does not know the actual field configurations, 
the probability for such an effective ``collaps" of the wave functional 
is exactly equal to the usual quantum mechanical probability 
for such a ``collaps". This is why our theory has the same statistical 
predictions as the usual theory. 
In the case in which all the effectivities are smaller than 1, 
which corresponds to a situation in which the wave functional 
has not ``collapsed" into a state with a definite number of 
particles, our theory is neither in agreement nor in contradiction 
with the standard theory. This is why the effectivity is a hidden 
variable. This is completely analogous to the Bohmian particle positions,
which agree with the standard quantum mechanical predictions only 
when the wave function effectively ``collapses" into a state with a 
definite particle position, while in other cases it is neither in 
agreement nor in contradiction with standard QM.
%

Thus our approach explains why detectors detect integer number 
of bosonic particles. There are also other attempts to explain this 
in the framework of causal interpretation 
of QFT \cite{bohmPR2,kal}, but these attempts do not incorporate 
particle trajectories. Here we repeat that there are also 
stochastic approaches to explain this 
for all fields \cite{bell1,bell2,durr} and a deterministic approach 
for fermions \cite{col}.  
%

Finally, it is fair to note that our approach explains why 
detectors detect integer number of particles only if an ideal 
measurement, based on nonoverlaping wave functionals, 
is assumed. One might consider this as a serious problem 
and conclude that a stochastic approach 
\cite{bell1,bell2,durr} better explains integer numbers of 
bosonic particles. However, we note that this problem
is analogous to a problem with the Bohmian interpretation 
of a nonrelativistic particle in a harmonic potential, which will 
be found to have energy equal to a
Hamiltonian eigenvalue $\omega(n+1/2)$
(with $n$ being an integer) only if the energy is measured through an 
ideal measurement (see the Appendix for a general argument). 
If a theory of quantum 
measurements is not taken into account, then Bohmian mechanics 
leads to statistical predictions that agree with the conventional 
quantum-mechanical predictions only when the statistical 
predictions refer to the {\em preferred} observables.
The preferred Bohmian observables are particle positions in the case of 
nonrelativistic QM and field configurations 
in the case of bosonic QFT. 
Particle momenta and energy in nonrelativistic QM 
and number of particles in bosonic QFT are not preferred observables in 
Bohmian mechanics, so the explanation of the conventional 
quantum-mechanical rules for statistical distributions of these 
observables requires a theory of quantum measurements.

\section{DISCUSSION}
\label{sec4} 

The results of this paper offer a  
solution to the problems of the current version
of the dBB interpretation of relativistic bosonic 
QM and QFT. We believe that they provide a 
deterministic interpretation of {\em all} physical effects
of the conventional bosonic QFT, including a deterministic interpretation
of the processes of creation and destruction of particles.
We have explicitly presented equations for real spin-0 fields, but 
the generalization to complex fields and other integer spins is 
straightforward. In particular,
particles and antiparticles resulting from a complex field $\phi$
possess the separate particle currents $j_{\mu}^{(P)}$ and 
$j_{\mu}^{(A)}$,
respectively, such that the usual charge current is
$j_{\mu}=i\phi^*\apm\phi=j_{\mu}^{(P)}-j_{\mu}^{(A)}$
\cite{nikol_PLB,nikol_long,nikol}. This means that particles and
antiparticles should be treated as different particle species, 
even when electromagnetic interactions are present \cite{nikol_long}.
Refs.~\cite{nikol_long,nikol} also contain the 
generalization of (\ref{loc2}) to spin-$1/2$ fields, so it is 
also straightforward to generalize the results of Sec.~\ref{sec2} 
to fermionic particles and antiparticles. 
However, it is not 
trivial to generalize the results of Sec.~\ref{sec3}  
to anticommuting fermionic fields, so this will be 
discussed in a separate paper.

It is also fair to note that the theory proposed in this paper 
may not be the only consistent 
solution to the problems of the current version
of the dBB interpretation of relativistic bosonic
QM and QFT. Moreover, it is possible that
some parts of the theory will turn out to be  
inconsistent, which will 
require further modifications of the theory. 

For example, 
we propose that both particles and fields  
objectively exist and that the macroscopic objects 
are made of both, but the question of consistency of such a picture 
requires further research. 
It is still possible that only particles or only 
fields should be fundamental objects in a dBB-type theory. 

Also, although (\ref{traj3l}) seems to us to be the most natural 
generalization of (\ref{traj3}), other generalizations are 
also possible. In general, particles moving 
according to (\ref{traj3l}) do not need to be distributed 
according to $j_0$ even if they were distributed 
so initially. However, 
this is not in contradiction with the conventional QFT 
simply because the conventional relativistic QFT,
in general, does not 
make clear probability predictions for distributions 
of particle positions. Therefore, our theory is able to 
give testable predictions on phenomena on which the conventional theory 
is not able to do that.

The remark above on nonexistence of predictions for distributions
of particle positions in QFT requires additional explanations.
In Refs.~\cite{bell1,bell2,col}, the operator of fermion-number 
density $\psi^{\dagger}\psi$ is defined, which, in turn, 
leads to predictions for distributions
of particle positions in fermionic QFT.
These predictions do not need to 
agree with the predictions that result from our theory,
which might be considered as a problem for our theory.
However, we do not consider this as a serious problem 
because, contrary 
to the claim in Refs.~\cite{bell1,bell2}, we do {\em not} consider 
the interpretation of the operator $\psi^{\dagger}\psi$ above
as a part of the {\em conventional} interpretation of 
fermionic QFT. Instead, by the conventional interpretation
(see, e.g., Ref.~\cite{ryder}) we understand the interpretation 
based on taking the normal ordering of the product 
$\psi^{\dagger}\psi$. This, owing to the anticommutative nature
of fermionic fields, leads to an operator with both positive and 
negative eigenvalues. Such normal-ordered operator is interpreted 
as the operator of charge density, which, in turn, leads to 
predictions for distributions of charge, not of particles.
For example, if the probability of finding charge at some point 
is equal to zero, it tells us nothing about the probability 
of finding particle-antiparticle pairs at that point. 
Similarly, for complex bosonic fields, the operator
$i\phi^{\dagger}\apnul\phi$ defines the charge density, not the 
particle density. For more details on the difference 
and similarities between the charge density and the particle 
density in QFT, see Refs.~\cite{nikol_PLB,nikol_long,nikol}.

The conventional QFT has definite predictions on angular 
distributions of particles produced 
in a scattering process, assuming that 
the particles are found in states with definite momenta. 
This assumption corresponds to wave functions of the form 
of (\ref{jed17}). In such a case, our theory predicts 
that particles move according to classical 
trajectories (straight lines) determined by their momenta, 
so the predictions on angular distributions are identical 
to those of the conventional QFT. Similarly, in the 
nonrelativistic limit without quantum field interactions, our 
theory of particle trajectories reduces to the usual 
dBB interpretation of QM, for which it is already known 
that it is in agreement with the predictions of the 
conventional nonrelativistic QM. Therefore, as far as we 
can see, all definite predictions of the 
conventional theory are also the predictions of our theory, 
provided that measurements are based on ideal quantum measurements 
(see the Appendix).

Note also that different definitions of the effectivity, 
replacing the definition (\ref{en}), are conceivable.
However, as already mentioned, the definition (\ref{en})
corresponds to a theory in which the probability 
of the particle existence in a stochastic interpretation
is equal to the effectivity of particles 
in a deterministic interpretation.
This leads to an appealing ontological picture in which the 
probability of existence is reinterpreted as  
a kind of ``degree of existence" (called effectivity)
which is {\em not} a probabilistic quantity. 
Finally, note that if all physical meaning of the quantity 
$e_n$ is given by Eq.~(\ref{goldstein}), then 
this equation explains precisely enough the physical meaning
of this quantity, so in this case 
it is not really necessary to use a funny name 
for it, such as ``effectivity". However, we view Eq.~(\ref{goldstein}) 
only as the simplest example of a possible precise meaning
of $e_n$. Different realizations of the general idea that (in 
some way) the effective mass is equal to the product $e_n m$
are conceivable. Therefore, to keep in mind the possibility of 
different realizations of the general abstract idea of a notion 
of effectivity, we retain the name ``effectivity" for the 
quantity $e_n$.  

The theory presented in this paper is certainly opened for 
further modifications, refinements, and reinterpretations.
We hope that new ideas introduced in this paper,
such as the abstract (and perhaps still somewhat vague) notion of 
effectivity, will motivate further research. 

\vspace{0.4cm}
\noindent
{\bf Acknowledgements.}
The author is grateful to S.~Goldstein and R.~Tumulka for their 
critical remarks and suggestions. In particular, the basic 
idea for Eq.~(\ref{goldstein}) arose from a suggestion 
of S.~Goldstein. The author is also grateful to anonymous 
referees for their constructive critical objections that stimulated
a more clear presentation. 
This work was supported by the Ministry of Science and Technology of the
Republic of Croatia under Contract No.~0098002.

\section*{APPENDIX: \\ THE GENERAL THEORY OF QUANTUM MEASUREMENTS}

In our experience, many physicists familiar with the 
dBB interpretation of QM are not familiar
with the corresponding general theory of quantum measurements, 
despite the fact that this theory is explained in often cited 
works on the dBB interpretation,
such as Refs.~\cite{bohm,bohmPR1,holbook}. More abstract presentations 
of this theory can also be found in Refs.~\cite{dur2,dur3}.
Since this theory of quantum measurements is also important 
for understanding of the present 
paper, in this Appendix we present a short review of this 
theory. For simplicity, we present this theory for the case 
of nonrelativistic QM with only one 
measured degree of freedom, but the same ideas can be easily adjusted 
to other quantum theories as well. Although this Appendix 
is intended to be self-contained, we note that an interested
reader can find more details in the references cited above.

Let $x$ denote the coordinate of particle position in the 
configuration space. Let $\psi(x,t)$ be the 
corresponding wave function. The dBB theory of particle motion 
is based on the postulate that the velocity $v$ of the particle 
is given by 
\begin{equation}
v=\frac{1}{m}\partial_x S,
\end{equation}
where $S(x,t)$ is the phase of the wave function. This postulate
provides that the statistical distribution of particle positions
is given by $|\psi(x,t)|^2$ for any time $t$, provided 
that this distribution is given by $|\psi(x,t_0)|^2$ 
for some initial time $t_0$. However, this fact by itself 
{\em is not sufficient} to provide the agreement of the 
dBB interpretation with the standard interpretation of QM. 
A simple way to see this is to consider the statistical 
distribution of momenta, which, according to the dBB interpretation, 
is
\begin{equation}\label{app2}
\rho(p,t)=\int dx\, |\psi(x,t)|^2 \delta(p-\partial_x S(x,t)).
\end{equation}
In general, this distribution is {\em not} equal to the 
quantum mechanical distribution $|\tilde{\psi}(p,t)|^2$ 
[where $\tilde{\psi}(p,t)$ is the Fourier transform of 
$\psi(x,t)$]. In order to see how the dBB interpretation 
recovers all the statistical results of standard QM, it is 
necessary to understand the general theory of quantum measurements.

Any measurement eventually reduces to an observation of some 
macroscopic quantity of the measuring apparatus, 
such as the position of a needle. In fact, this macroscopic observation 
can allways be eventually reduced to an observation of the 
{\em position} (in the configuration space) of something. Let us 
idealize and simplify the analysis by introducing only one 
configuration-space variable $y$ 
corresponding to the measuring apparatus. (Introducing a larger 
number of such variables does not change the conclusions.)
Assume that we want to construct a measuring apparatus that 
measures an observable represented by a hermitian operator $\hat{A}$
that acts on the $x$ space. 
The wave function $\psi(x,t)$ can be expanded as 
\begin{equation}
\psi(x,t)=\sum_a c_a(t)\psi_a(x),
\end{equation}
where $\psi_a(x)$ are complete normalized eigenfunctions 
of the operator $\hat{A}$:
\begin{equation}
\hat{A}\psi_a(x)=a\psi_a(x).    
\end{equation}
For simplicity, we assume that the spectrum of the eigenvalues $a$ 
is not degenerate.
According to standard QM, the probability of finding the state 
to have the value $a$ of the observable $\hat{A}$ is equal 
to $|c_a(t)|^2$. On the other hand, Eq.~(\ref{app2}) suggests that 
this may not be the case in the dBB interpretation, 
unless $\hat{A}$ is equal to $x$.
To see how this problem resolves, it is essential to realize
that, when the system consisting of the variables $x$ and $y$ can be 
considered as a configuration suitable for measurement of the 
$x$ subsystem, the total wave function is {\em not} of the form 
$\psi(x,t)\chi(y,t)$. Instead, the interaction 
between the $x$ subsystem and the $y$ subsystem should be such that 
the total wave function takes the form
\begin{equation}\label{app4}
\Psi(x,y,t)=\sum_a c_a(t)\psi_a(x)\chi_a(y),
\end{equation}
where the normalized wave functions $\chi_a(y)$ 
with different labels $a$ do not overlap in the $y$ space. 
Therefore, if the position $y$ is found to have the value in the 
support of a wave function $\chi_a(y)$, then this value is not in 
the support of any other wave function $\chi_{a'}(y)$. 
In other words, if the position $y$ is found to have the value in the 
support of $\chi_a(y)$, then, according to the usual 
rules of QM, we know that the total wave function is, effectively, 
equal to $\psi_a(x)\chi_a(y)$. The probability for this to happen is, 
according to (\ref{app4}), equal to $|c_a(t)|^2$, just as it should 
be without taking into account any theory of quantum measurements.

The discussion of the preceding paragraph is valid without 
taking into account the dBB interpretation of QM. However, 
without the dBB interpretation, it is not clear why and how the variable 
$y$ takes a definite value. On the other hand, if the variable $y$ 
is also described by the dBB interpretation, then it becomes clear 
why and how it takes a definite value. 
From the Bohmian mechanics 
of composed systems and the fact that the wave functions $\chi_a(y)$
do not overlap, it is easy to show that if the particle describing 
the measuring apparatus has the position $y$ in the support of 
$\chi_a(y)$, then the measured particle with the 
position $x$ moves in the same way 
as it was described by the wave function $\psi_a(x)\chi_a(y)$. 
In such a case, 
the value of the observable $\hat{A}$ is a constant of motion  
equal to $a$. In this way, the wave function $\Psi(x,y,t)$ 
effectively ``collapses" into a wave function $\psi_a(x)\chi_a(y)$, 
by $y$ taking a definite value from the support of $\chi_a(y)$.
In the same sense, the wave function $\psi(x,t)$ 
effectively ``collapses" into a wave function $\psi_a(x)$. 
It only remains to see that the probability for this to happen 
is equal to $|c_a(t)|^2$. We know that, in the dBB interpretation, 
the probability density in the configuration space is 
\begin{equation}
|\Psi(x,y,t)|^2=\sum_a |c_a(t)|^2 |\psi_a(x)|^2 |\chi_a(y)|^2, 
\end{equation}
where the fact that different $\chi_a(y)$ do not overlap has been used,
which has eliminated the nondiagonal terms proportional to 
$\chi_a\chi_{a'}=0$ for $a\neq a'$. By averaging over $x$, we find
the probability distribution in the $y$ space to be 
\begin{equation}
\rho(y,t)=\sum_a |c_a(t)|^2 |\chi_a(y)|^2=|\chi(y,t)|^2,
\end{equation}
where
\begin{equation}
\chi(y,t)\equiv \sum_a c_a(t) \chi_a(y).
\end{equation}
This shows that the dBB interpretation predicts that the probability 
for $y$ to take a value from the support of $\chi_a(y)$ is 
equal to $|c_a(t)|^2$.


\begin{thebibliography}{99}

\bibitem{bohm}
D.~Bohm, {\it Phys.~Rev.}~{\bf 85}, 166, 180 (1952).
\bibitem{bohmPR1}
D.~Bohm and B.~J.~Hiley, 
{\it Phys. Rep.}~{\bf 144}, 323 (1987).
\bibitem{bohmPR2}
D.~Bohm, B.~J.~Hiley, and P.~N.~Kaloyerou,
{\it Phys.~Rep.}~{\bf 144}, 349 (1987).
\bibitem{holPR}
P.~R.~Holland, {\it Phys.~Rep.}~{\bf 224}, 95 (1993).
\bibitem{holbook}
P.~R.~Holland, {\it The Quantum Theory of Motion}
(Cambridge University Press, Cambridge, 1993).
\bibitem{bohmPRL}
D.~Bohm and B.~J.~Hiley, {\it Phys.~Rev.~Lett.}~{\bf 55}, 2511 (1985).
\bibitem{bohmN}
D.~Bohm, C.~Dewdney, and B.~J.~Hiley, {\it Nature} {\bf 315}, 294 (1985).
\bibitem{dewN}
C.~Dewdney, P.~R.~Holland, A.~Kyprianidis, and J.~P. Vigier,
{\it Nature} {\bf 336}, 536 (1988).
\bibitem{cuf}
N.~Cufaro-Petroni, C.~Dewdney, P.~Holland, T.~Kyprianidis,
and J.~P.~Vigier, {\it Phys.~Lett.~A} {\bf 106}, 368 (1984).
\bibitem{kyp}
A.~Kyprianidis, {\it Phys.~Lett.~A} {\bf 111}, 111 (1985).
\bibitem{dew}
C.~Dewdney, G.~Horton, M.~M.~Lam, Z.~Malik, and M.~Schmidt,
{\it Found.~Phys.}~{\bf 22}, 1217 (1992).
\bibitem{holNC}
P.~R.~Holland and J.~P.~Vigier, {\it Nuovo Cimento} {\bf 88B}, 20 (1985).
\bibitem{hor}
G.~Horton, C.~Dewdney, and A.~Nesteruk, {\it J.~Phys.~A} 
{\bf 33}, 7337 (2000).
\bibitem{holfer}
P.~R.~Holland, {\it Phys.~Lett.~A} {\bf 128}, 9 (1988).
\bibitem{sud}
A.~Sudbery, {\it J.~Phys.~A} {\bf 20}, 1743 (1987).
\bibitem{durr}
D.~D\"urr, S.~Goldstein, R.~Tumulka, and N.~Zanghi, 
{\it J.~Phys.~A} {\bf 36}, 4143 (2003).
\bibitem{shoj}
A.~Shojai and M.~Golshani, quant-ph/9612023.
\bibitem{col}
S.~Colin, {\it Phys.~Lett.~A} {\bf 317}, 349 (2003).
\bibitem{bell1}
J.~S.~Bell, {\it Phys.~Rep.}~{\bf 137}, 49 (1986).
\bibitem{bell2}
J.~S.~Bell, {\it Speakable and Unspeakable in Quantum
Mechanics}
(Cambridge University Press, Cambridge, 1987).
\bibitem{nikol_PLB}
H.~Nikoli\'c, {\it Phys.~Lett.~B} {\bf 527}, 119 (2002);
Erratum {\bf 529}, 265 (2002).
\bibitem{nikol_long}
H.~Nikoli\'c, {\it Int.~J.~Mod. Phys.~D} {\bf 12}, 407 (2003).
\bibitem{nikol}
H.~Nikoli\'c, hep-th/0205022.
\bibitem{cost}
J.~P.~Costella, B.~H.~J.~McKellar, and A.~A.~Rawlinson,
{\it Am.~J.~Phys.}~{\bf 65}, 835 (1997).
\bibitem{bd}
N.~D.~Birrell and P.~C.~W.~Davies, {\it Quantum Fields in
Curved Space} (Cambridge Press, NY, 1982).
\bibitem{long}
D.~V.~Long and G.~M.~Shore, {\it Nucl.~Phys.~B} {\bf 530}, 247 (1998).
\bibitem{schweber}
S.~S.~Schweber, {\it An Introduction to Relativistic Quantum Field Theory}
(Harper \& Row, New York, 1961).
\bibitem{kal}
P.~N.~Kaloyerou, quant-ph/0311035.
\bibitem{ryder}
L.~H.~Ryder, {\it Quantum Field Theory} (Cambridge University Press, 
Cambridge, 1984). 
\bibitem{dur2}
D.~D\"urr, S.~Goldstein, and N.~Zanghi,
{\it J.~Stat.~Phys.}~{\bf 67}, 843 (1992).
\bibitem{dur3}
D.~D\"urr, S.~Goldstein, and N.~Zanghi,            
quant-ph/0308038.

\end{thebibliography}
\end{document}